\documentclass[conference]{IEEEtran}
\IEEEoverridecommandlockouts
\usepackage{cite}

\ifCLASSINFOpdf
   \usepackage[pdftex]{graphicx}
   \DeclareGraphicsExtensions{.pdf,.jpeg,.png}
\else
  \usepackage[dvips]{graphicx}
  \DeclareGraphicsExtensions{.eps}
\fi

\graphicspath{{./images/}} 

\usepackage{textcomp}
\usepackage{xcolor}

\usepackage{epstopdf}
\usepackage{amsthm}
\usepackage{lipsum}
\usepackage{color}
\usepackage{algorithm}
\usepackage{algorithmic}
\usepackage{enumerate}

\usepackage{etoolbox}  
\makeatletter
\patchcmd{\algorithmic}{\addtolength{\ALC@tlm}{\leftmargin} }{\addtolength{\ALC@tlm}{\leftmargin}}{}{}
\usepackage{amssymb}

\ifCLASSINFOpdf
   \usepackage[pdftex]{graphicx}
\else
\fi
\usepackage{cite}
\usepackage[cmex10]{amsmath}
\usepackage{algorithmic}
\usepackage{array}
\usepackage{url}
\usepackage{makeidx}
\usepackage{verbatim}
\usepackage{subfigure}
\usepackage{multirow}
\newcommand{\beqn}{\begin{eqnarray}}
\newcommand{\eeqn}{\end{eqnarray}}
\newcommand{\bse}{\begin{subequations}}
\newcommand{\ese}{\end{subequations}}

\usepackage{algorithm}
\usepackage{algorithmic}

\usepackage{etoolbox}  
\makeatletter
\patchcmd{\algorithmic}{\addtolength{\ALC@tlm}{\leftmargin} }{\addtolength{\ALC@tlm}{\leftmargin}}{}{}
\makeatother

\makeatletter 
\newcommand{\linebreakand}{%
  \end{@IEEEauthorhalign}
  \hfill\mbox{}\par
  \mbox{}\hfill\begin{@IEEEauthorhalign}
}
\makeatother 

\begin{document}

\title{Evaluate Quantum Combinatorial Optimization for Distribution Network Reconfiguration
}

\author{\IEEEauthorblockN{Phuong Ngo, Christan Thomas, Hieu Nguyen, and Abdullah Eroglu}
\IEEEauthorblockA{{Dept. of Electrical \& Computer Eng., North Carolina A\&T State University}\\
{\{ango1, cqthomas\}@aggies.ncat.edu, \{htnguyen1, aeroglu\}@ncat.edu}}
\and
\IEEEauthorblockN{Konstantinos Oikonomou}
\IEEEauthorblockA{{Pacific Northwest National Laboratory}\\
konstantinos.oikonomou@pnnl.gov}}

%
%
%
%

\maketitle

\begin{abstract}
This paper aims to implement and evaluate the performance of quantum computing on solving combinatorial optimization problems arising from the operations of the power grid. 
To this end, we construct a novel mixed integer conic programming formulation for the reconfiguration of radial distribution network in response to faults in distribution lines.
Comparing to existing bus injection model in the literature, our formulation based the branch flows model is theoretically equivalent without needing non-explainable variables, thus being more numerically stable.
The network reconfiguration model is then used as a benchmark to evaluate the performance of quantum computing algorithms in real quantum computers. 
It shows that while current quantum computing algorithms with fast execution time in quantum computers can be a promising solution candidate, its heuristic nature stem from its theoretical foundation should be considered carefully when applying into power grid optimization problems.

\end{abstract}

\begin{IEEEkeywords}
network reconfiguration, quantum computing, alternating direction method of multipliers, quantum approximation optimization algorithm.
\end{IEEEkeywords}

{ 
\footnotesize
\section*{Nomenclature}
\addcontentsline{toc}{section}{Nomenclature}
\subsection{Set and Indices}
\begin{IEEEdescription}[\IEEEusemathlabelsep\IEEEsetlabelwidth{$V_1,V_2,V_3$}]
\raggedright 
\item[$\mathcal{I}, i$]  Set and index of distribution bus, $i \in \mathcal{I}$
\item[$\ell =ij$] Distribution line connecting bus $i$ to bus $j$
\item[$\mathcal{L}$]  Set of distribution lines, $\ell =ij \in \mathcal{L}$
\item[$\mathcal{SW}$] Set of lines which have a tie switch
\end{IEEEdescription}

\subsection{Parameters}
\begin{IEEEdescription}[\IEEEusemathlabelsep\IEEEsetlabelwidth{$V_1,V_2,V_3$}]
\raggedright 
\item[$r_\ell$,$x_\ell$] Resistance/reactance of line $\ell$
\item[$\overline{V}_{i}$, $\underline{V}_{i}$] Maximum/minimum voltage magnitude at bus $i$
\item[$P_{i}^d, Q_{i}^d$] Real/reactive power demands at bus $i$
\item[$VOLL$] Value of lost load of customer
\item[$\delta_{\ell}$] Binary parameter, set to 0 if line $\ell$ is outage, otherwise 1
\end{IEEEdescription}

\subsection{Variables}
\begin{IEEEdescription}[\IEEEusemathlabelsep\IEEEsetlabelwidth{$V_1,V_2,V_3$}]
\raggedright
\item[$P_{ij}$] Real power flow from bus $i$ to bus $j$
\item[$Q_{ij}$] Reactive power flow from bus $i$ to bus $j$
\item[$P^{\sf grid}_{}$] Total real power imported to the main grid
\item[$Q^{\sf grid}_{}$] Total reactive power imported to the main grid
\item[$\nu_{i}$] Square voltage at node $i$
\item[$\nu_{i}^\ell$] Auxiliary variable associated with squared voltage at node $i$ on line $\ell$
\item[$I_{\ell}^{sq}$] Square current on line $\ell$
\item[$\alpha_{\ell}$] Binary variable represented to switching action, set to 1 if line $\ell$ is closed, otherwise 0
\item[$\beta_{ij}$] Variable represented to a parent-child relationship between node $i$ and node $j$, set to 1 if node $i$ is a child of node $j$, otherwise 0
\item[$P_{i}^{ds}$] Real power load demand served at bus $i$
\item[$P_{i}^c$] Curtailed real power load demand at bus $i$
\item[$Q_{i}^{ds}$] Reactive power load demand served at bus $i$
\item[$Q_{i}^c$] Curtailed reactive power load demand at bus $i$
\item[$R_{\ell},T_{\ell}$] Auxiliary variables associated with line $\ell$ to formulate the conic constraint
\end{IEEEdescription}
}

\section{Introduction}
There are tremendous advantages in quantum computing in recent years.
The size of quantum computers have been over 100 qubits, resulting in significant computational capabilities in comparison to classical computers.
There are also significant works on mathematical foundation and algorithm designs leveraging the quantum processors. 
Applications of quantum computing for power systems have been also recently investigated, i.e.,   solving DC power flow  \cite{eskandarpour2021experimental} and fast-decouple AC power equations \cite{feng2021quantum}.
However, since these problems can tackled in poly-nominal times with high accuracy in classical computers, quantum computing can be expected to improve the computational time without solution performance concern.

Many problems in power grids are NP-hard  combinatorial optimization problems, where both computational time and solution accuracy are of concerns.
There are several efforts to extend the quantum computing algorithms into NP-hard problems, such as combinatorial optimization, utilizing the qubits' representation of binary variables \cite{qaoa, qaoa_performance}.  
While there are some promising results, it is worth noting that the mathematical foundation of these algorithms are heuristic. 
Therefore, examining the performance of these algorithms on power system applications is needed, especially to inform quantum  researchers on specific solution designs for power grid. 

To this end, this paper constructs combinatorial optimization problems arising from distribution network reconfiguration.
These include two mixed integer conic programs, extended from bus injection and branch flows models \cite{Jabr2012reconfi, Lowc2013convex}, for fault isolation and network reconfiguration.
Our models, which can effectively model the line flows in connected and disconnected lines without using Big-M formulation as in \cite{ShanShan2018-NR, Mehdi2020-NR}, can be used as a benchmark for evaluating existing quantum combinatorial algorithms \cite{qaoa}. 
The paper is organized as follows.
Section II presents and compares two
equivalent formulations of network reconfiguration, which are mixed integer conic programs. 
 Section III introduces basics
of quantum physics and state-of-the-art tools for quantum
optimization. Then, we present how we can tailor quantum
computing algorithms to solve the network reconfiguration
problem. 
Numerical results are presented in section IV, which indicate the heuristic nature of existing quantum algorithms for combinatorial optimization problem might need to be considered when applying to power system domains.
Section V concludes the paper.

\section{Network Reconfiguration}

\begin{figure}[t!]
    \centering
	\includegraphics[width=0.7\linewidth]{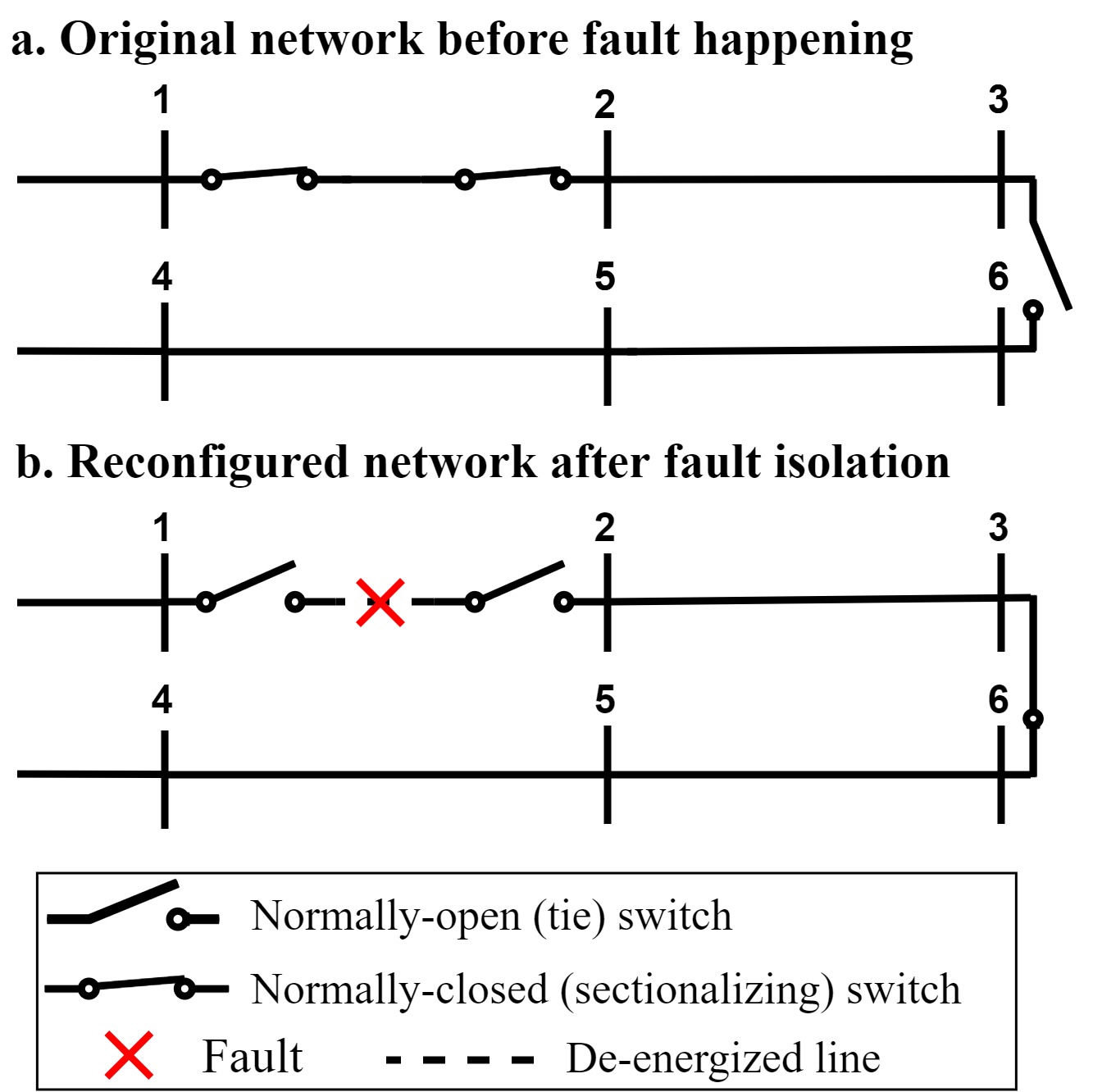}
	\vspace{-0.2pt}
	\caption{Fault Isolation and Restoration Procedure}
	\label{fig: FISR}
\end{figure}
A radial power distribution network can be represented as a connected graph with no loops $\mathcal{G} = \langle \mathcal{I}, \mathcal{L} \rangle$, 
where $\mathcal{I}$ and $\mathcal{L}$ are sets of distribution nodes and lines. 
$\mathcal{L}$ includes normally-closed lines and tie switches (normally-open).
Once a fault occurs in a line, the normally-closed sectionalizing switches in the faulted line will open to isolate the fault. Then, normally-open tie switches close  to re-energize the healthy parts of the network.
This reconfiguration procedure as shown in Fig. \ref{fig: FISR} is to minimize out-of-service loads respecting the network radiality structure.
Let $\delta_\ell$ represent status of the normally-closed lines, either connected or disconnected due to fault and $\alpha_\ell$ represent the switching status (close or open) of lines with tie-switches.  
If the line $\ell$ is connected, (i,e,, no fault or tie switch is closed) the line power flows are:
\begin{subequations}
\label{linepowerflow}
    \beqn
        	P_{ij}= \frac{r_{ij}}{(r_{ij}^2+x_{ij}^2)}V_i^2 - \frac{r_{ij}}{(r_{ij}^2+x_{ij}^2)} V_i V_j \cos\theta_{ij} \label{Rab_Pij_1} \notag \\
        	- \frac{r_{ij}}{(r_{ij}^2+x_{ij}^2)} V_i V_j \sin\theta_{ij}\\
        	Q_{ij}= - \frac{x_{ij}}{(r_{ij}^2+x_{ij}^2)}V_i^2 + \frac{x_{ij}}{(r_{ij}^2+x_{ij}^2)} V_i V_j \cos\theta_{ij} \label{Rab_Qij_1} \notag \\
        	- \frac{x_{ij}}{(r_{ij}^2+x_{ij}^2)} V_i V_j \sin\theta_{ij}, \\
        	~~\textrm{if}~~  \delta_\ell =1~ \textrm{and/or} ~ \alpha_\ell =1 \nonumber
    \eeqn
where $P_{ij}$ and $Q_{ij}$ be the real and reactive powers that come out of node $i$ and are sent to node $j$ over the line $\ell = ij$ with impedance $z_{ij} = r_{ij} + j x_{ij}$ as shown in Figure \ref{fig: Impedance}. 
Also, ${\mathop V\limits^\to}_i= V_{i} \angle \theta_i$ and ${\mathop V\limits^\to}_j = V_{j} \angle \theta_j$ are nodal voltages at $i$ and $j$  where $V$ and $\theta$ represent voltage magnitude and  angle respectively.
\begin{figure}[http]
    \centering
	\includegraphics[width=0.7\linewidth]{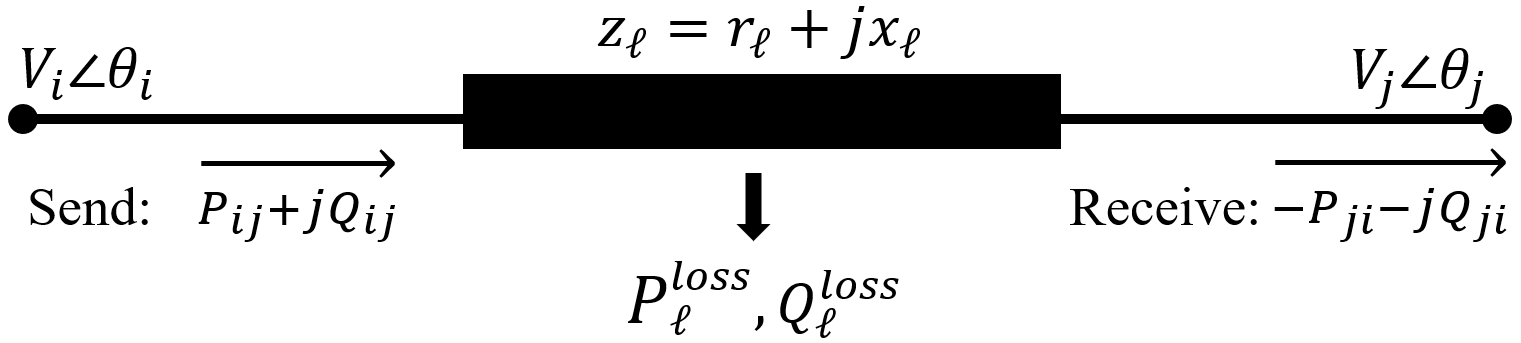}
	\vspace{-0.2pt}
	\caption{Power flow on a distribution line $\ell = ij$}
	\label{fig: Impedance}       
\end{figure}    
If the line is disconnected, i.e., either by a fault ($\delta_\ell =0$) or opening the tie switch ($\alpha_\ell =0$), the line flows should be  zero: 
\beqn
P_{ij} =0 ,  Q_{ij} =0 ~~\textrm{if}~~  \delta_\ell =0~ \textrm{or} ~ \alpha_\ell =0
\eeqn 
\end{subequations}
In general, the network reconfiguration model must model line flow equation conditioning on the connection status of the line (\ref{linepowerflow}) respecting the network radiality structure.
Generally, this is a mixed-binary nonlinear optimization due to a set of binary variables representing the switching actions and to the non-convex nature of power flow constraints.
However, in the case of radial network and under mild assumption, the problem can be formulated as mixed integer conic program (MICP), which can be solved optimally by commercial solvers such as CPLEX and MOSEK in classical computers.
Therefore, it can be an ideal benchmark for evaluating new computing methods in power systems, particularly the emerging trend of quantum computing.  
This paper first develops novel MICP formulations for radial distribution network reconfiguration to examine the performance of quantum combinatorial optimization algorithms. 
Specifically, we consider two formulations, one is based on existing bus injection model \cite{Jabr2012reconfi}, which employs non-physical auxiliary variables, and the other developed in this paper, which is extended from the branch flows model \cite{Lowc2013convex}.

\subsection{The bus injection based reconfiguration model}

The bus injection model \cite{Jabr2012reconfi} used two auxiliary variables:
\beqn
\label{rab_equality}
    R_{\ell} = V_i V_j \cos\theta_{ij},~ T_{\ell} = V_i V_j \sin\theta_{ij}, ~~\forall \ell =ij \in \mathcal{L}
\eeqn
which yields the following conditions:
\beqn
    (R_{\ell})^2 + (T_{\ell})^2 = (V_i)^2 (V_j)^2 , ~ R_{\ell}  \geq 0,  \quad \forall \ell = ij \in \mathcal{L} \label{Rab_RotatedCone}.
\eeqn
Then, the network configuration is as follows:

\bse
	\label{jabr}
\beqn
    \min \Bigg( P^{\sf grid} + VOLL \sum_{i=2}^n P^c_i \Bigg). \label{objrab}
\eeqn
\ese
Subject to:
\begin{itemize}

 \item{Radiality constraints}
\bse
    \beqn
    	\beta_{1j}=0, \quad j \in \mathcal{N}(1) \label{beta_v1}\\
        \beta_{ij} \in \left[0,1\right], \quad \forall i \in \mathcal{I}, ~j\in \mathcal{N}(i) \label{beta_v2}\\
        \sum_{j\in \mathcal{N}(i)} \beta_{ij}=1, \quad \forall i \in \mathcal{I} \label{beta_v3}\\
        \beta_{ij} + \beta_{ji} = \alpha_{\ell}, \quad \forall \ell = ij \in \mathcal{SW} \label{beta_v4}\\
        \beta_{ij} + \beta_{ji} = \delta_{\ell}, \quad \forall \ell = ij \in \mathcal{L} \backslash \mathcal{SW} \label{beta_v5}
    \eeqn
\ese

 \item{Incorporating line connection statuses}
\bse
    \beqn
        0 \leq \nu_{i}^\ell \leq (\overline{V_i})^2 \alpha_{\ell} \label{alpha_v1}, ~~ \forall \ell=ij \in \mathcal{SW}\\
        0 \leq \nu_{i} - \nu_{i}^\ell \leq (\overline{V_i})^2(1 - \alpha_{\ell}) \label{alpha_v2}, ~~ \forall \ell=ij \in \mathcal{SW}\\
        0 \leq \nu_{j}^\ell \leq (\overline{V_j})^2 \alpha_{\ell} \label{alpha_v3}, ~~ \forall \ell=ij \in \mathcal{SW}\\
        0 \leq \nu_{j} - \nu_{j}^\ell \leq (\overline{V_j})^2(1 - \alpha_{\ell}) \label{alpha_v4}, ~~ \forall \ell=ij \in \mathcal{SW}\\
        0 \leq \nu_{i}^\ell \leq (\overline{V_i})^2 \delta_{\ell} \label{delta_v1}, ~~ \forall \ell=ij \in \mathcal{L} \backslash \mathcal{SW}\\
        0 \leq \nu_{i} - \nu_{i}^\ell \leq (\overline{V_i})^2(1 - \delta_{\ell}) \label{delta_v2}, ~~ \forall \ell=ij \in \mathcal{L} \backslash \mathcal{SW}\\
        0 \leq \nu_{j}^\ell \leq (\overline{V_j})^2 \delta_{\ell} \label{delta_v3}, ~~ \forall \ell=ij \in \mathcal{L} \backslash \mathcal{SW}\\
        0 \leq \nu_{j} - \nu_{j}^\ell \leq (\overline{V_j})^2(1 - \delta_{\ell}) \label{delta_v4}, ~~ \forall \ell=ij \in \mathcal{L} \backslash \mathcal{SW}
    \eeqn
\ese
\item{Line power flow constraints}
\bse
\beqn
			(R_{\ell})^2 + (T_{\ell})^2 \leq \nu_i^\ell \nu_j^\ell \label{Rab_convex}, ~ R_{\ell}  \geq 0,  \quad \forall \ell = ij \in \mathcal{L}\\
			P_{ij} = \frac{1}{r^2_{\ell} + x^2_\ell} \left( r_{\ell} \nu_i^\ell - r_\ell R_{\ell} + x_\ell T_{\ell} \right), \label{Rab_Pij} \forall \ell = ij \in \mathcal{L}\\
			Q_{ij} = \frac{1}{r^2_{\ell} + x^2_\ell} \left( x_{\ell} \nu_i^\ell - x_\ell R_{\ell} - x_\ell T_{\ell} \right) , \label{Rab_Qij} \forall \ell = ij \in \mathcal{L}\\
			P_{ji} = \frac{1}{r^2_{\ell} + x^2_\ell} \left( r_{\ell} \nu_j^\ell - r_\ell R_{\ell} - x_\ell T_{\ell} \right) \label{Rab_Pji}, \forall \ell = ij \in \mathcal{L} \\
			Q_{ji} = \frac{1}{r^2_{\ell} + x^2_\ell} \left( x_{\ell} \nu_j^\ell - x_\ell R_{\ell} + x_\ell T_{\ell} \right) \label{Rab_Qji}, \forall \ell = ij \in \mathcal{L}\\
			I^{sq}_{\ell} = \frac{1}{r^2_{\ell} + x^2_\ell} \left( \nu_i^\ell + \nu_j^\ell - 2R_{\ell} \right) \geq 0 \label{Rab_Isq},  \forall \ell = ij \in \mathcal{L},
\eeqn
\ese

 \item {Operational constraints based on the bus injection model}
 \bse
		\beqn
	        (\underline{V}_i)^2 \leq \nu_i \leq (\overline{V}_i)^2, \label{Rab_vollim} \quad \forall i \in \mathcal{I}\\
			P^{d}_i = P^{ds}_i + P^{c}_i \label{Rab_Prelax}, \quad \forall i \in \mathcal{I} \\
    	    Q^{d}_i = Q^{ds}_i + Q^{c}_i \label{Rab_Qrelax}, \quad \forall i \in \mathcal{I} \\
			P^{ds}_i, P^{c}_i, Q^{ds}_i, Q^{c}_i \geq 0, ~~ \forall i \in \mathcal{I} \label{Rab_Nonneg}\\
			P^{ds}_{i} + \sum_{j \in \mathcal{N} \left(i \right)} P_{ij} = 0 \label{Rab_Pbalance}, \quad \forall i \in \mathcal{I} \\
			Q^{ds}_{i} + \sum_{j \in \mathcal{N} \left( i \right)} Q_{ij} = 0 \label{Rab_Qbalance}, \quad \forall i \in \mathcal{I} \\
			P^{\sf grid} = \sum\limits_{j \in \mathcal{N}(1)} P_{1j},  ~
            Q^{\sf grid} = \sum\limits_{j \in \mathcal{N}(1)}} {Q_{1j} \label{Rab_SlackPQ}.
		\eeqn
	\ese   
\end{itemize}

The objective function  is to maximize the amount customers' demand that can be served in the network by reconfigurating the network in response to the fault, in other words, minimize total quantity of curtailed power with penalty VOLL.

The set of radiality constraints \eqref{beta_v1}-\eqref{beta_v4} aim to maintain the radial structure of the network. 
Here, we use the parent-child relation-based method, which is also called the spanning tree condition \cite{Jabr2012reconfi}. 
It regulates every single node, apart from the substation node, to have only one parent expressed in \eqref{beta_v1} and \eqref{beta_v2}, where $\beta_{ij}$ and $\beta_{ji}$ represent a relation between node $i$ and $j$. 
If $\beta_{ij}$ equals to 1, node $i$ is parent of node $j$ and 0 otherwise.
Here, we model $\beta_{ij}$ as continuous variables constrained in $\left[0,  1\right]$, but they only take value of 0 or 1 as proved in Claim 1 of \cite{taylor2012convex}. 
Therefore, the set of radiality constraints in this paper save a number of $(|\mathcal{I}|^2 - 2 \times \mathcal{L})$ binary variables compared to the original formulation in \cite{Jabr2012reconfi}.

In order to model the power flows over the connected and disconnected lines, we need to incorporate the line connection statuses in the computation of the power flows.
To this end, for each line $\ell$ we define the two auxiliary variables $\nu^\ell_i$ and $\nu^\ell_j$ such that

$$
	\begin{cases}
		\nu^\ell_i  = \nu_i, ~\nu^\ell_j  = \nu_j, & \text{if $\ell$ is connected}\\
		\nu^\ell_i  = \nu^\ell_j  = 0, & \text{if $\ell$ is disconnected} \label{connection_status}
	\end{cases}
$$
where $\nu_i, \nu_j$ represent the squared nodal voltages. 
The connection status condition (\ref{connection_status}) is captured by 
constraints \eqref{alpha_v1}-\eqref{delta_v4} where the connection is based on the values of $\delta_\ell$ and $\alpha_\ell$. 

The set of real and real active power flows are shown in (\ref{Rab_convex})-(\ref{Rab_Isq}). 
Obviously, when line is connected, i.e., $\delta_\ell =1$ or $\alpha_\ell =1$, the $\nu^\ell_i = \nu_i$ and $\nu^\ell_j = \nu_j$ and (\ref{Rab_Pij})-(\ref{Rab_Isq}) indeed represent the original line flows and line current formulation as in (\ref{linepowerflow}). 
On the other hand, if the line is disconnected, i.e., $\delta_\ell =0$ or $\alpha_\ell =0$, the $\nu_i^\ell, \nu_j^\ell$ are forced to be 0 because of \eqref{alpha_v1}-\eqref{delta_v4}.
Consequently, $R_\ell =0$ because of (\ref{Rab_convex}).
Then, the values of $P_{ij}, Q_{ij}, P_{ji}$ and $Q_{ji}$ calculated in (\ref{Rab_Pij})-(\ref{Rab_Qji}) are also zero.
Additionally, the line current $I^{sq}_\ell$ also equal to zero. 

Finally, constraint \eqref{Rab_vollim} represents the nodal voltage limits,  real and reactive power load served at each node are modeled in  \eqref{Low_Prelax}-\eqref{Low_Qrelax} where $P^{ds}_i, Q^{ds}_i$ and $P^c_i, Q^c_i$ are the amount of demand served and curtailment at node $i$, and should be positive variables as in (\ref{Rab_Nonneg}).
Additionally, real and reactive power balances at each node is given in (\ref{Rab_Pbalance} and (\ref{Rab_Qbalance}). 

The overall problem is an MICP where the rotated cone (\ref{Rab_convex}) is the convex relaxation of the original equality condition (\ref{rab_equality}), which is generally exact for radial networks. 
However, such the formulation contains non-explainable variables $R_\ell$ and $T_\ell$ that do not hold any physical meaning.
Also, as discussed in \cite{KPEC_ACOPF}, the rotated cone (\ref{Rab_convex}) can exhibit a numerical ill-condition due to the very small gap $|\nu_i^\ell - \nu_j^\ell|$ arising when MICP solvers transform it into the equivalent quadratic cone. 


\subsection{Distribution network reconfiguration formulation based on the branch flows model}

We aim to develop a more compact and numerically stable  formulation of network configuration without needing non-explainable auxiliary variables by extending the branch flows model \cite{Lowc2013convex} with line connection status. 
The model is as follows:

\bse
    \label{Low}
\begin{equation}
    \min \Bigg( P^{\sf grid} + VOLL \sum_{i=2}^n P^c_i \Bigg) \label{objLow}
\end{equation}

Subject to:
\begin{itemize}
    \item{Radiality constraints}
    \eqref{beta_v1}-\eqref{beta_v4}

    \item{Incorporating line connection statuses}
    \eqref{alpha_v1}-\eqref{delta_v4}
    
    \item{Operational constraints based on the branch flows model}
\beqn
		P_{ij}^2 + Q_{ij}^2 \leq \nu_i^\ell I^{sq}_{\ell}, \label{Low_convex}~~ \forall\ell=ij\in \mathcal{L}\\
		\nu_j^\ell = \nu_i^\ell - 2(r_{\ell} P_{ij} + x_{\ell} Q_{ij})+(r_{\ell}^2 + x_{\ell}^2)I^{sq}_{\ell}, \label{Low_Isq} \notag \\
		\forall ij=\ell \in \mathcal{L}\\
    	(\underline{V}_{i})^2 \leq \nu_i \leq (\overline{V}_{i})^2 \label{Low_Vollim}, \quad \forall i \in \mathcal{I} \\
    	P^{d}_i = P^{ds}_i + P^{c}_i \label{Low_Prelax}, \quad \forall i \in \mathcal{I} \\
    	Q^{d}_i = Q^{ds}_i + Q^{c}_i \label{Low_Qrelax}, \quad \forall i \in \mathcal{I} \\
		\sum_{k: k \to i}\big(P_{ki} - r_{ki} I^{sq}_{ki} \big) - \sum_{j: i \to j } P_{ij} = P^{ds}_{i} \label{Low_Pij}, ~~ \forall i \in \mathcal{I} \\
		\sum_{k: k \to i}\big(Q_{ki} - x_{ki} I^{sq}_{ki} \big) - \sum_{j: i \to j } Q_{ij} = Q^{ds}_{i} \label{Low_Qij}, ~~ \forall j \in \mathcal{I} \\
		P^{\sf grid} = \sum\limits_{j: 1 \to j} P_{1j},~~ Q^{\sf grid} = \sum\limits_{j: 1\to j } Q_{1j} \label{Low_SlackPQ} \\
		I^{sq}_{\ell} \geq 0,  \quad \forall \ell \in \mathcal{L}, \nonumber\\
		~ P^{ds}_i, P^{c}_i, Q^{ds}_i, Q^{c}_i \geq 0, ~~ \forall i \in \mathcal{I}. \label{Low_Nonneg}
    \eeqn

\end{itemize}    
  \ese

The objective function, radiality constraints, and line connection constraints are the same with the previous bus injection model.
Operational constrains (\ref{Low_convex})-\eqref{Low_Nonneg} are constructed following the branch flows model. 
Here, the second order cone relaxation is constructed from the circular relation between real and reactive line powers $P_{ij}, Q_{ij}$. 
When lines are connected, these operational constraints model the original power flow equations over the distribution lines as in \cite{Lowc2013convex}.
When line $\ell= ij$ are disconnected, the values of $\nu^\ell_i, \nu^\ell_j$ are forced to zero.
Consequently, both $P_{ij}, Q_{ij}$ are forced to zero in accordance to (\ref{Low_convex}).
Also, the voltage drop equation (\ref{Low_Isq}) enforces $I^{sq}_\ell$ to be zero, i.e., there is no current over the disconnected line. 

\subsection{Related fault isolation and network reconfiguration model}



Compare to the bus injection model (\ref{objrab})-(\ref{Rab_SlackPQ}), the extended branch flows model exhibits a better numerical stability arising from different representation of the rotated cones as discussed in \cite{KPEC_ACOPF}.
Compare other models in the literature \cite{ShanShan2018-NR, Mehdi2020-NR}, the our model does not need Big-M term to decouple the line power flows and nodal voltage in the disconnected lines. 
Choosing the value of Big-M  is not a trivial task.
If M is chosen to be too large, it leads to an instability of numerical results. 
If M is chosen to be too small, it might fail to properly model the decoupling of physical variables, i.e., line flows and voltages,  in disconnected lines. 



\section{Quantum Computing based Solution Approaches}
This section discusses the backgrounds of quantum computing and how we leverage quantum combinatorial optimization algorithms to solve the network reconfiguration problem, which is an MICP, in a real quantum computer. 

\subsection{Quantum Computing Basics}
Quantum computing has logic operations based on the principles of quantum mechanics, which differs from classical computing. 
The fundamental unit of information in quantum computation is the qubit. 
The state of a qubit has a two-dimensional complex vector space illustrated as follows:
\[
| \psi \rangle = a | 0 \rangle + b | 1 \rangle : a, b \in \mathbb{C} \wedge |a|^2 + |b|^2 = 1.
\]
The coefficients $a$ and $b$ are complex numbers known as probability amplitudes that satisfy the constraint $|a|^2 + |b|^2 = 1$.
A single qubit can be in a superposition of two basis states $|0\rangle$ and $|1\rangle$.
Therefore, a system of $n$-qubits contains an arbitrary superposition of $2^n$ basis states \cite{BurVerify_Qiskit_Circuit}.
Moreover, every single one of the $2^n$ super-positioned computations reserves the information in the same binary variable.
For instance, if an optimization task would occupy $2^7 = 128$ bits to be executed in a classical computer, a quantum computer would be about to only need 7 qubits.
Therefore, it is expected that a polynomial-time of a quantum computing machine is more powerful than that of a classical computing machine in certain search algorithms.
The recent development of quantum computing platforms, services, development kits such as IBM Quantum, Azure Quantum, AWS Quantum, D-Wave Systems, IonQ, Q\# and Qiskit, has a strong motivation to a speed-up from quantum computing solutions applied to combinatorial optimization problems.
However, existing quantum computing algorithms, particularly for solving combinatorial optimization problems, employ relaxation of binary constraints using quantum states.
As indicated in their mathematical foundation paper  \cite{qaoa} is heuristic, thus it is worth examine its performance on combinatorial optimization problems arising from power system operations, e.g., distribution network reconfiguration.  

\begin{figure}[http]
    \centering
	\includegraphics[width=0.8\linewidth]{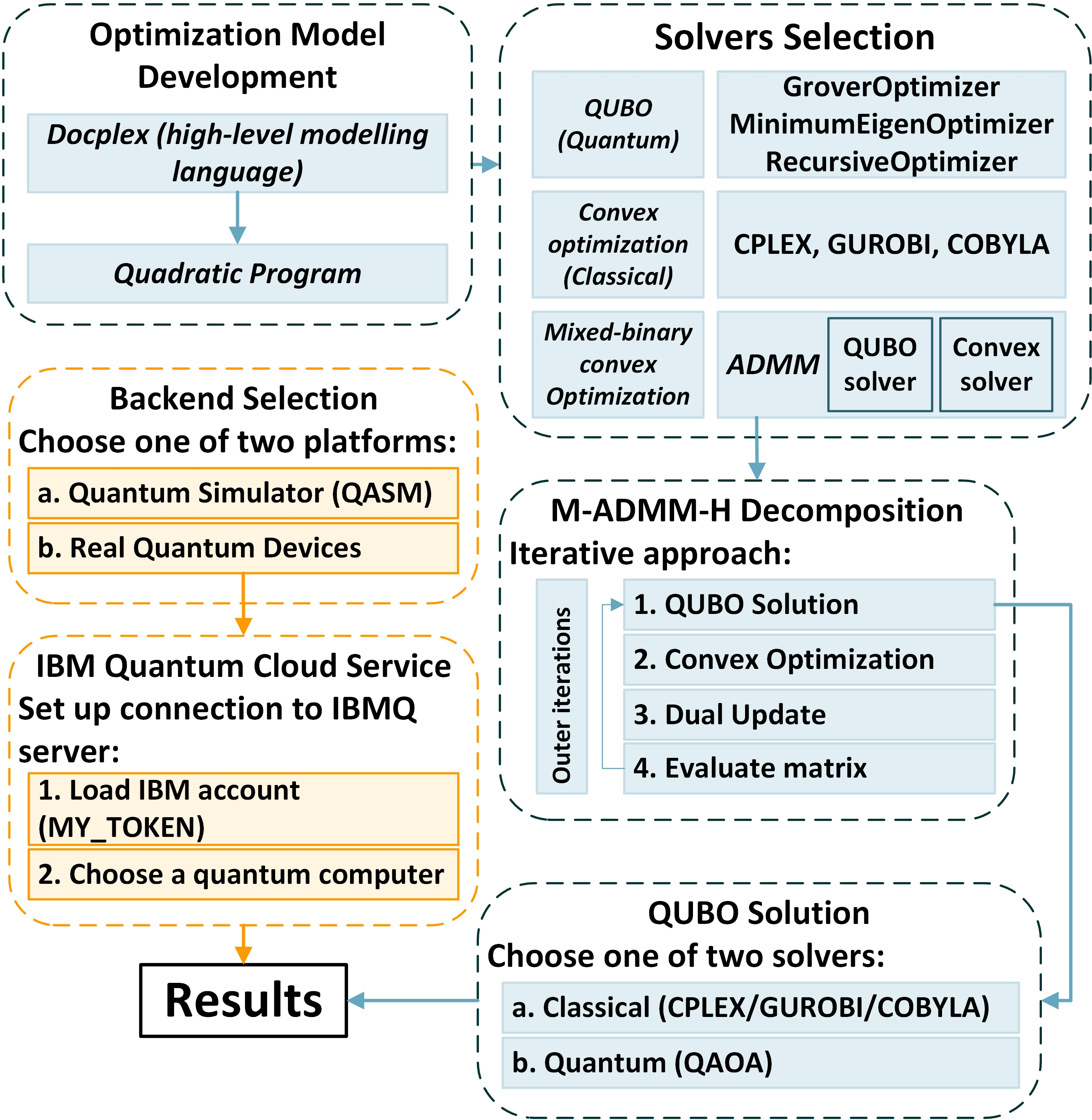}
	\vspace{-0.2pt}
	\caption{Solution Techniques Workflow}
	\label{fig: Qiskit_Workflow}
\end{figure}

\subsection{Workflows for working with quantum computers}
In this paper, we use Qiskit - an open-source SDK \cite{qiskit} to test the network reconfiguration formulation with quantum computing enabled by IBM Quantum services \cite{ibmq}.
The overall work flow is shown in in Fig. \ref{fig: Qiskit_Workflow}
First, the MICP based network reconfiguration problem should be built in appropriate syntax that can be compatible with quantum optimization algorithms.
There are two methods, i.e., direct and indirect ones.
In direct method, a program can be coded directly following structures and syntaxes regulated by Qiskit. 
In indirect method, we formulate the MICP program in conventional mathematical modelling languages such as docplex or gurobi, which will be converted into a “quadratic program". 
Here, “quadratic program" in quantum computing refer the optimization problem contains quadratic constraints and some binary variables, each binary variable can be approximated by quantum state. 
If the program is successfully converted or built, an appropriate solver should be selected based on the characteristics of program. 
If the program only contains binary variables, particularly, purely integer program in classical optimization, GroverOptimizer,  MinimumEigneOptimizer or RecursiveOptimizer should be chosen.
If the program also contains continuous variables, i.e., mixed-binary optimization program as in the network reconfiguration problem, the ADMMOptimizer should be selected.

\subsection{Alternating Direction Method of Multipliers for Convex Optimization}


The alternating direction method of multipliers (ADMM)-based hybrid quantum-classical computing (QCC) algorithm is used to solve a class of mixed-binary convex optimization problems \cite{ADMM}. 
The MICP is decomposed in into a quadratic unconstrained binary optimization (QUBO) problem containing only binary variables, and convex sub-problems, which only contain continuous variables where the relaxed binary variables are fixed.
The QUBO problem is solved by  quantum approximate optimization algorithm (QAOA) by relaxing binary by quantum states whereas
the convex problems is solved by conventional solvers such as CPLEX, GUROBI, and COBYLA.
The accuracy of results by quantum computing relies on the effect of inexact result of the binary sub-problem \cite{ADMM} and noise of real quantum computers \cite{Quantum_noise_2022}.
Consequently, ADMM can not guarantee the convergence for optimal solution due to the non-smooth nature of mixed-binary problem.

\subsection{Quantum Approximate Optimization Algorithm}
The quantum approximate optimization algorithm (QAOA) is particularly appropriate to problems containing only binary variables
Specifically, the $N$ binary variables in the original integer program are considered in a form of $N$-bit strings: $x_1,x_2,...,x_N$. 
QAOA converts the original integer optimization problem into the searching for optimal decision variables in a form of bitstring such that the following optimization ratio is maximized \cite{qaoa}:
\[
    R = \frac{f(x)}{\max_x (f(x)}, \label{ratio}
\]
where $f(x)$ denotes the objective of the original integer program and R is called constant-gap approximation.
Obviously, $0 \leq R \leq 1$. 
However, $\max_x (f(x)$ in (\ref{ratio}) is the optimal values with optimal solutions of the original integer optimization we have to find, thus being unknown for the algorithm.
On the other hand, we know the hardness of binary approximation ratio $R$, (e.g., the limit of approximation for the MaxCut problem is 16/17$\approx$0.94117) \cite{arora1998proof}. 
Therefore, the key point is to find $x$ such that the ratio $f(x)/f(\hat{x})$ improves and closes to its theoretical limits ($\hat{x}$ is the round values of the relaxed $x \in [0,1]$ back to $\{0,1\}$ domain). 

In QAOA, the searching procedure utilizes the quantum state relaxation of binary variables. 
Obviously, it is a heuristic algorithm \cite{qaoa}. 
However, QAOA is one of leading potential algorithms as it takes advantage of quantum computing power in finding solution for optimization problems \cite{qaoa_performance}.


\




\section{Numerical Results}
\begin{figure}[t!]
    \centering
	\includegraphics[width=0.8\linewidth]{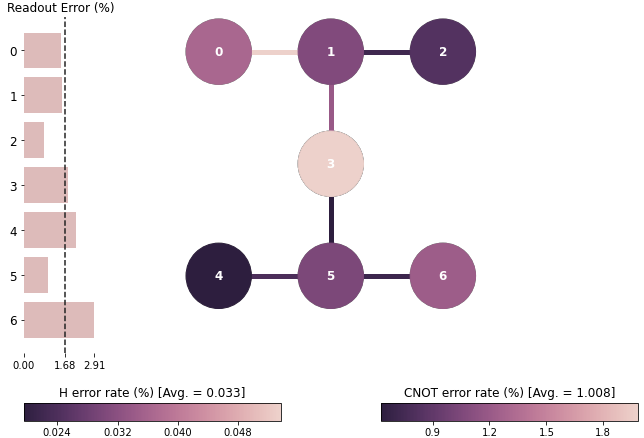}
	\vspace{-0.2pt}
	\caption{Coupling map of $ibm\_oslo$ server}
	\label{fig: Decouple_map}
\end{figure}

We examine three numerical performance of three approaches: (i) and (ii) solving the MICP based network reconfiguration (both the bus injection model and the developed branch flows model) on a classical computer using a classical optimizer, i.e., CPLEX and (iii) solving the branch flows based network configuration model in a real quantum computer.
Specifically, we use the $ibm\_oslo$, a 7-qubit machine provided by IBM depicted in Figure \ref{fig: Decouple_map}, to solve the constructed MICP.   

\begin{figure}[t!]
    \centering
	\includegraphics[width=0.85\linewidth]{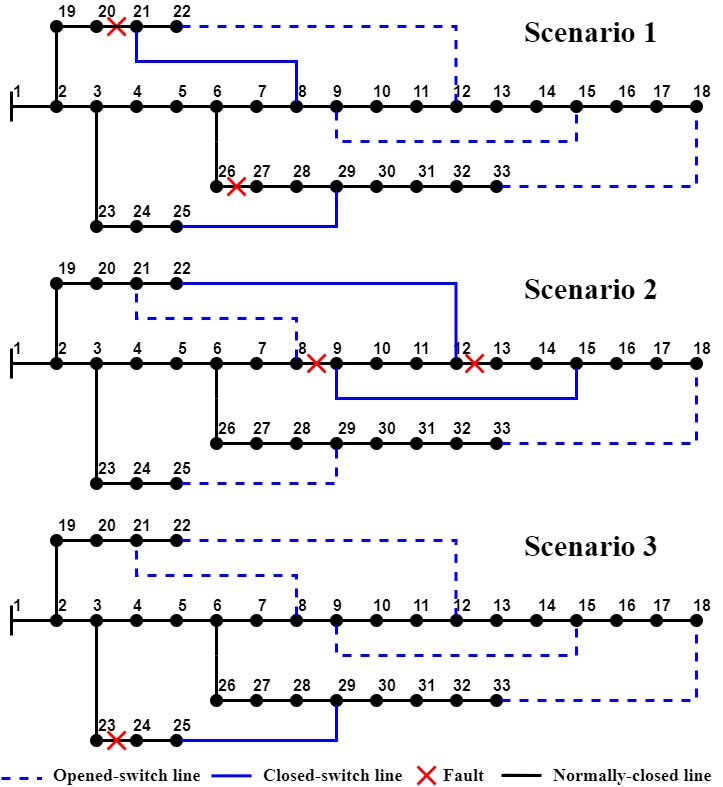}
	\vspace{-0.2pt}
	\caption{Fault scenarios and obtained network reconfiguration}
	\label{fig: Scenario}
\end{figure}

\begin{figure}[t!]
    \centering
	\includegraphics[width=0.95\linewidth]{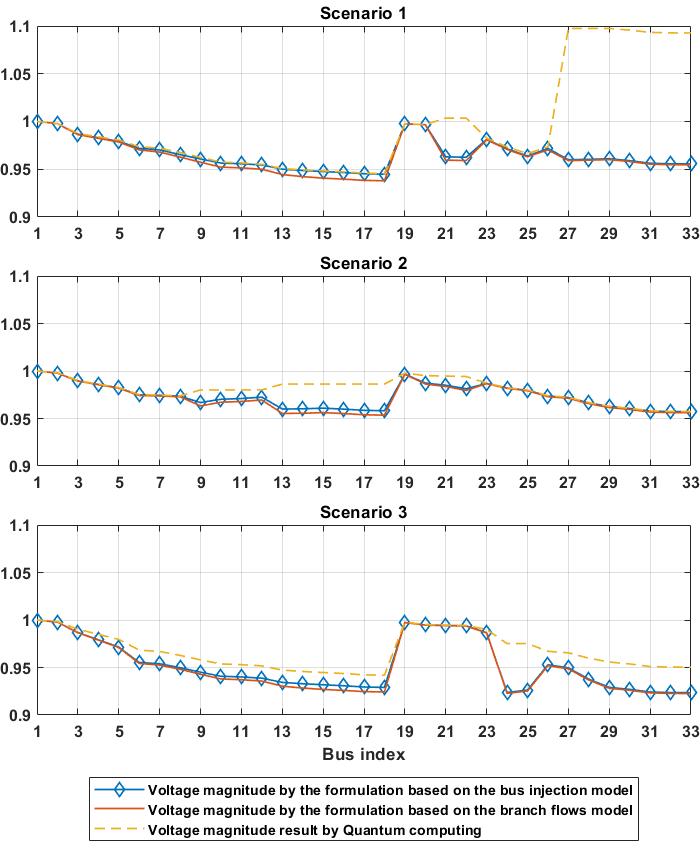}
	\vspace{-0.2pt}
	\caption{Voltage magnitude (in pu)}
	\label{fig: voltage}
\end{figure}

\begin{figure}[t!]
    \centering
	\includegraphics[width=0.95\linewidth]{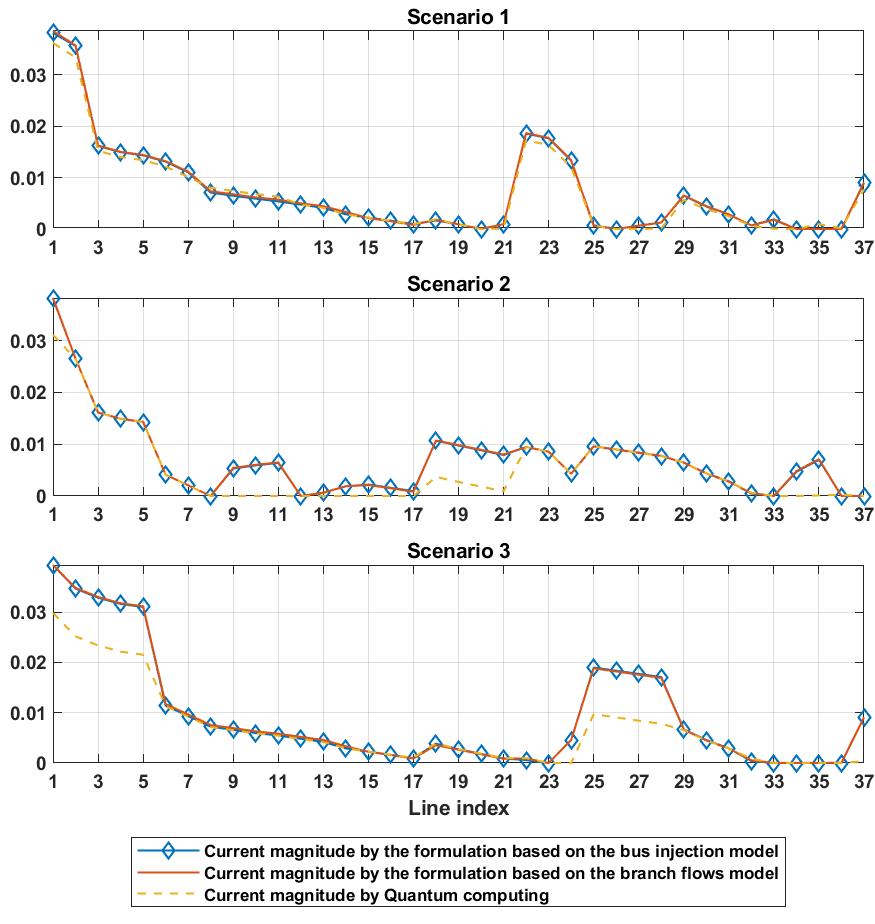}
	\vspace{-0.2pt}
	\caption{Current magnitude (in pu)}
	\label{fig: current}
\end{figure}

We conduct the numerical results on the IEEE 33 nodes considering three random scenarios of faults on the network as shown in Fig. \ref{fig: Scenario}.
The three methods result in the same switching actions, i.e., exact binary of classical solvers, and quantum results after rounding into binary values.  
Fig. \ref{fig: voltage} and \ref{fig: current} show that the voltage magnitude and current magnitude results of both formulations on the classical PC are equivalent with insignificant differences.
On the other hand, the quantum computing solutions exhibit some differences from the real optimal ones.
This can be explained by the heuristic nature of the algorithms.
Another reason is the large number of inequalities in the problem, which have to be converted into equalities resulting in difficult convergence  \cite{qaoa}. 
Additionally, the accuracy of final quantum computing solution relies heavily on the solution of QUBO and the noise of quantum device.

\section{Conclusions}

This paper presents two mixed integer conic program formulation for the fault isolation and network reconfiguration problem, namely the existing  bus injection and new branch flows models. 
The branch flows-based network reconfiguration does not require non-explainable variables or Big-M term to model the line flows over connected and disconnected lines.
Its compact and numerical stable form can be used to evaluate new computing methods proposed for power system operations.
Our experiment with the only known combinatorial optimization algorithm on quantum computers, namely ADMM-QCC, shows that the quantum computing can be a promising approach for large-scale problems.
However, its results is not as accurate as by classical optimizers on classical computers, because of its heuristic nature. 
Therefore, the application of quantum computing on critical operations of power grid requires more investigation considering the mathematical foundation behind quantum algorithms.



\bibliographystyle{IEEEtran}
\bibliography{NAPS_Q}

\begin{thebibliography}{10}
\providecommand{\url}[1]{#1}
\csname url@samestyle\endcsname
\providecommand{\newblock}{\relax}
\providecommand{\bibinfo}[2]{#2}
\providecommand{\BIBentrySTDinterwordspacing}{\spaceskip=0pt\relax}
\providecommand{\BIBentryALTinterwordstretchfactor}{4}
\providecommand{\BIBentryALTinterwordspacing}{\spaceskip=\fontdimen2\font plus
\BIBentryALTinterwordstretchfactor\fontdimen3\font minus
  \fontdimen4\font\relax}
\providecommand{\BIBforeignlanguage}[2]{{%
\expandafter\ifx\csname l@#1\endcsname\relax
\typeout{** WARNING: IEEEtran.bst: No hyphenation pattern has been}%
\typeout{** loaded for the language `#1'. Using the pattern for}%
\typeout{** the default language instead.}%
\else
\language=\csname l@#1\endcsname
\fi
#2}}
\providecommand{\BIBdecl}{\relax}
\BIBdecl

\bibitem{eskandarpour2021experimental}
R.~Eskandarpour, K.~Ghosh, A.~Khodaei, and A.~Paaso, ``Experimental quantum
  computing to solve network dc power flow problem,'' \emph{arXiv preprint
  arXiv:2106.12032}, 2021.

\bibitem{feng2021quantum}
F.~Feng, Y.~Zhou, and P.~Zhang, ``Quantum power flow,'' \emph{IEEE Transactions
  on Power Systems}, vol.~36, no.~4, pp. 3810--3812, 2021.

\bibitem{qaoa}
E.~Farhi, J.~Goldstone, and S.~Gutmann, ``A quantum approximate optimization
  algorithm,'' 2014.

\bibitem{qaoa_performance}
G.~E. Crooks, ``Performance of the quantum approximate optimization algorithm
  on the maximum cut problem,'' \emph{arXiv:Quantum Physics}, 2018.

\bibitem{Jabr2012reconfi}
R.~A. Jabr, R.~Singh, and B.~C. Pal, ``Minimum loss network reconfiguration
  using mixed-integer convex programming,'' \emph{IEEE Trans. Power Syst.},
  vol.~27, no.~2, pp. 1106--1115, May 2012.

\bibitem{Lowc2013convex}
M.~Farivar and S.~H. Low, ``Branch flow model: Relaxations and
  convexification{\textemdash}part i,'' \emph{IEEE Trans. Power Syst.},
  vol.~28, no.~3, pp. 2554--2564, 2013.

\bibitem{ShanShan2018-NR}
A.~Arif, S.~Ma, and Z.~Wang, ``Dynamic reconfiguration and fault isolation for
  a self-healing distribution system,'' in \emph{2018 {IEEE/PES} Transmission
  and Distribution Conference and Exposition ({T\&D})}.\hskip 1em plus 0.5em
  minus 0.4em\relax IEEE, Apr. 2018.

\bibitem{Mehdi2020-NR}
M.~M. Hosseini and M.~Parvania, ``Computationally efficient formulations for
  fault isolation and service restoration in distribution systems,'' in
  \emph{2020 {IEEE} Power \& Energy Society General Meeting ({PESGM})}.\hskip
  1em plus 0.5em minus 0.4em\relax IEEE, Aug. 2020.

\bibitem{taylor2012convex}
J.~A. Taylor and F.~S. Hover, ``Convex models of distribution system
  reconfiguration,'' \emph{IEEE Transactions on Power Systems}, vol.~27, no.~3,
  pp. 1407--1413, 2012.

\bibitem{KPEC_ACOPF}
A.~P. Ngo, C.~Thomas, K.~Oikonomou, H.~Nguyen, and D.~Nguyen, ``On the
  comparison of different convexified power flow models in radial network,'' in
  \emph{2022 {IEEE} Kansas Power and Energy Conference ({KPEC})}.\hskip 1em
  plus 0.5em minus 0.4em\relax IEEE, Apr. 2022.

\bibitem{BurVerify_Qiskit_Circuit}
L.~Burgholzer, R.~Raymond, and R.~Wille, ``Verifying results of the {IBM}
  qiskit quantum circuit compilation flow,'' in \emph{2020 {IEEE} International
  Conference on Quantum Computing and Engineering ({QCE})}.\hskip 1em plus
  0.5em minus 0.4em\relax IEEE, Oct. 2020.

\bibitem{qiskit}
\BIBentryALTinterwordspacing
{IBM Qiskit Development Team}, ``Qiskit: An open-source framework for quantum
  computing,'' 2022. [Online]. Available:
  \url{https://qiskit.org/documentation}
\BIBentrySTDinterwordspacing

\bibitem{ibmq}
\BIBentryALTinterwordspacing
{IBM Quantum team}, ``Ibm quantum,'' 2022. [Online]. Available:
  \url{https://quantum-computing.ibm.com/}
\BIBentrySTDinterwordspacing

\bibitem{ADMM}
C.~Gambella and A.~Simonetto, ``Multiblock {ADMM} heuristics for mixed-binary
  optimization on classical and quantum computers,'' \emph{IEEE Trans. Quantum
  Eng.}, vol.~1, pp. 1--22, 2020.

\bibitem{Quantum_noise_2022}
X.~Liu, A.~Angone, R.~Shaydulin, I.~Safro, Y.~Alexeev, and L.~Cincio, ``Layer
  {VQE}: A variational approach for combinatorial optimization on noisy quantum
  computers,'' \emph{IEEE Trans. Quantum Eng.}, vol.~3, pp. 1--20, 2022.

\bibitem{arora1998proof}
S.~Arora, C.~Lund, R.~Motwani, M.~Sudan, and M.~Szegedy, ``Proof verification
  and the hardness of approximation problems,'' \emph{Journal of the ACM
  (JACM)}, vol.~45, no.~3, pp. 501--555, 1998.

\end{thebibliography}

\end{document}